# A radiation tolerant clock generator for the CMS Endcap Timing Layer readout chip

H. Sun,[a,b,1] Q. Sun,[c] S. Biereigel,[d,e] R. Francisco,[d] D. Gong,[b,2] G. Huang,[a] X. Huang,[b] S. Kulis,[d] P. Leroux,[e] C. Liu,[b] T. Liu,[b] T. Liu,[c] P. Moreira,[d] J. Prinzie,[e] J. Wu,[c] J. Ye,[b] L. Zhang,[a,b,1] W. Zhang [a,b,1]

[a] *Central China Normal University,*
   *Wuhan, Hubei 430079, PR China*

[b] *Southern Methodist University,*
   *Dallas, TX 75275, USA*

[c] *Fermi National Accelerator Laboratory,*
   *Batavia, IL 60510, USA*

[d] *CERN,*
   *1211 Geneva 23, Switzerland*

[e] *ESAT-ADVISE research lab, KU Leuven University,*
   *3000 Leuven, Belgium*

E-mail: dgong@smu.edu

ABSTRACT: We present the test results of a low jitter Phase Locked Loop (PLL) prototype chip for the CMS Endcap Timing Layer readout chip (ETROC). This chip is based on the improved version of a clock synthesis circuit named ljCDR from the Low-Power Gigabit Transceiver (lpGBT) project. The ljCDR is tested in its PLL mode. An automatic frequency calibration (AFC) block with the Triple Modular Redundancy (TMR) register is developed for the LC-oscillator calibration. The chip was manufactured in a 65 nm CMOS process with 10 metal layers. The chip has been extensively tested, including Total Ionizing Dose (TID) testing up to 300 Mrad and Single Event Upset (SEU) testing with heavy ions possessing a Linear energy transfer (LET) from 1.3 to 62.5 $MeV \times cm^2/mg$.

---

[1] Visiting scholars at SMU and performed this work at SMU.
[2] Corresponding author.

# Contents

## 1. Introduction

A low-jitter and radiation-tolerant clock generator is a critical component for a front-end readout chip of a precision timing detector. The Minimum Ionizing Particles (MIP) Timing Detector (MTD) is a new detector planned for CMS during the High-Luminosity Large Hadron Collider (HL-LHC) era [1]. MTD consists of the Barrel Timing Layer (BTL) and Endcap Timing Layer (ETL), and the latter chooses Low Gain Avalanche Diodes (LGADs) as the sensors to deposit the charged particles' energy and generate charges through ionization. We have been developing the ETL readout chip (ETROC) based on a 65 nm CMOS process, aiming to measure the arrival time of impinging particles with a time resolution of 30~40 ps. ETROC requires a Phase Locked Loop (PLL) to provide precise and multi-frequency clocks (40 MHz, 320 MHz, 1.28 GHz, and 2.56 GHz) to the functional blocks within the chip, with the demand of the RMS jitter within 5 ps. This clock generator, named ETROC PLL, is required to survive 100 Mrad Total Ionizing Dose (TID) and be insensitive to Single Event Effects (SEEs).

    ETROC PLL adapted the improved version (January 2020) of ljCDR, a mature Clock and Data Recovery (CDR), and PLL circuit inside the Low-Power Gigabit Transceiver (lpGBT) project [2,3]. A low-noise and radiation-tolerant LC-tank Voltage-Controlled Oscillator (VCO) [4] was integrated with a nominal frequency of 5.12 GHz. A new charge pump in the PLL mode was used with improved static offset across different VCO capacitor configurations. Besides, the patterned ground shield was added under the inductor. ETROC PLL was developed in a 65 nm CMOS process with 10 metal layers and integrated into a standalone chip for characterization and change verification. In this work, the implementation and test results of the ETROC PLL prototype chip are discussed.

## 2. Circuit Design

### 2.1 Overall structure

The ETROC PLL core includes the ljCDR, the prescaler, the feedback divider from the lpGBT project, and an automatic frequency calibration (AFC) block used for LC-tank VCO. Figure 1 illustrates the block diagram of the ETROC PLL prototype chip. The ljCDR operates in PLL mode with a 40 MHz reference clock, and the CDR mode is disabled. The prescaler consists of the clock divider (N=2) and the level adapters from differential Current Mode Logic (CML) to single-ended CMOS signals. The feedback divider (N=64) generates clocks with proportional frequencies and employs Triple Modular Redundancy (TMR) to protect against Single Event Upsets (SEUs). Additional circuits include the input reference clock receiver, the output CML drivers, and a generic I$^2$C block. The reference voltage generator (1 V nominal), designed for the charge injection and the threshold voltage generator for each ETROC pixel, is also implemented. All power supply voltages, marked in different colors in the diagram, are separated to avoid power interferences and ease monitoring of power supply currents.

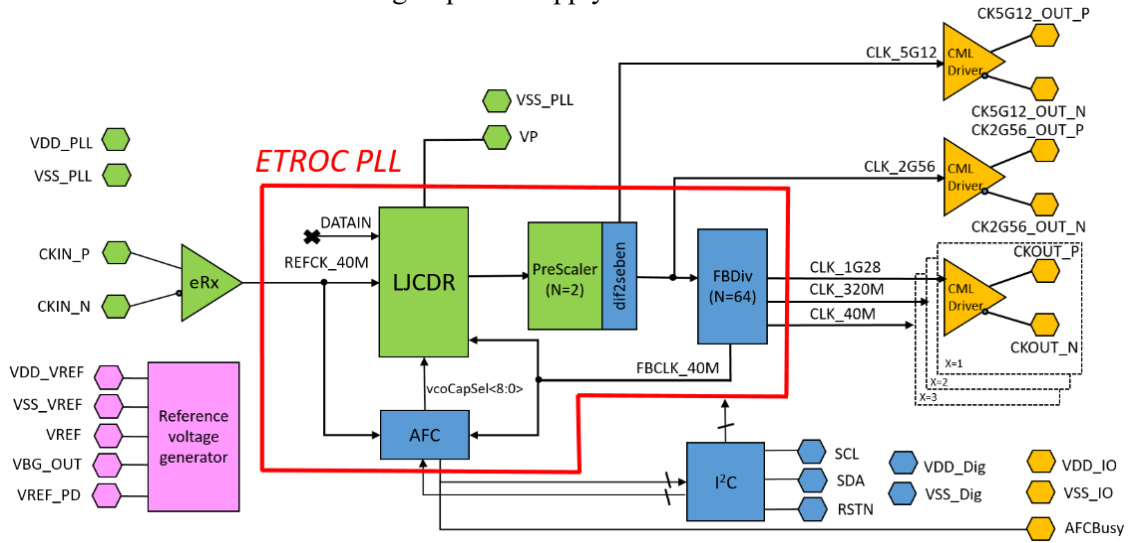

**Figure 1**: Block diagram of ETROC PLL standalone chip.

## 2.2 AFC

An AFC block is implemented to calibrate the LC-tank VCO in an optimal status automatically. The LC-tank VCO contains 8 banks of switched capacitors for a relatively large tuning range. For the expected oscillation frequency of 5.12 GHz, the capacitor bank should be carefully chosen and preset in advance to lock the PLL. The AFC block searches for the optimal capacitor bank when the tuning voltage is overridden. Since the VCO frequency decreases monotonically with switched capacitors, the binary search algorithm is applied to reduce the number of comparisons and speed up the calibration process. In the calibration process, the external reference clock and the feedback clock are divided by 4096 with 12-bit counters to compare their frequencies. In each comparison, the counters start counting after they are reset, and stop counting once a counter overflows. The carryout bit of the overflowed counter indicates whether the feedback clock is faster or slower than the reference clock. After the calibration process is complete, the calibrated data "Capsel" is stored and refreshed automatically in a TMR register to avoid data corruption due to SEUs. Figure 2(a) illustrates how AFC operates with the VCO and the slow control I$^2$C block. The fully automatic calibration workflow is shown in figure 2(b). After power-on, the PLL loop is turned on in the normal mode (default). Users then reset the AFC block and launch AFC calibration via the I$^2$C command. Once the calibration is finished, the monitor signal "AFCbusy" becomes low, and PLL returns to normal mode. If not, users can revise the loop parameters and restart the calibration until the PLL loop automatically settles.

## 3. Lab Test Results

The ETROC PLL, with the layout area of 1.2 mm × 0.7 mm, was implemented in a standalone test chip whose size is 2 mm × 1 mm. A photograph of the dedicated board is shown in figure 3(a), and Figure 3(b) illustrates how an ETROC PLL die was wire-bonded on the board. For potential laser testing, we drilled a hole on the test board below the chip with the same size of the possible sensitive area (the prescaler and the CML clock distribution) known from the lpGBT v0 testing.

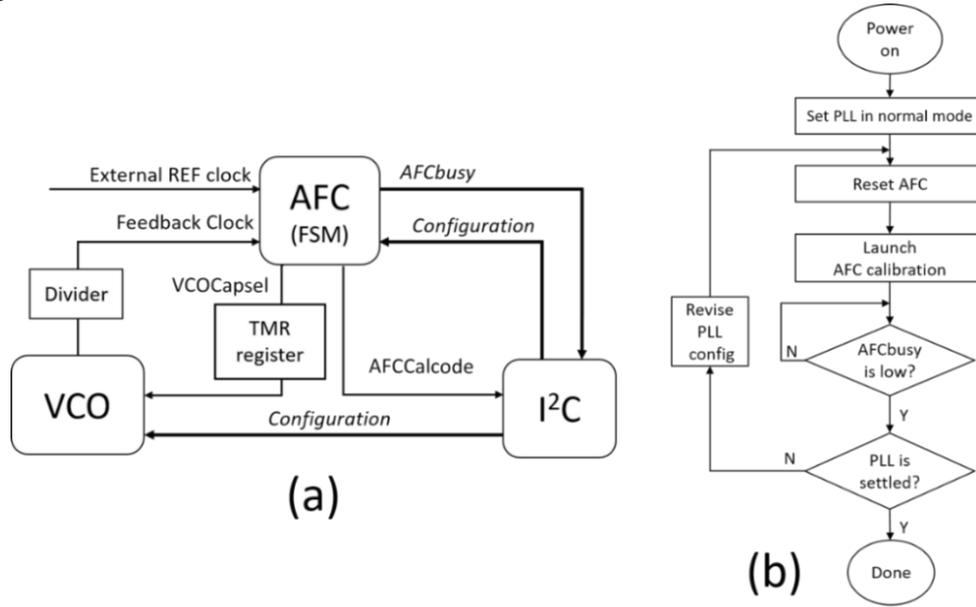

**Figure 2:** Block diagram (a) and workflow of the AFC block (b).

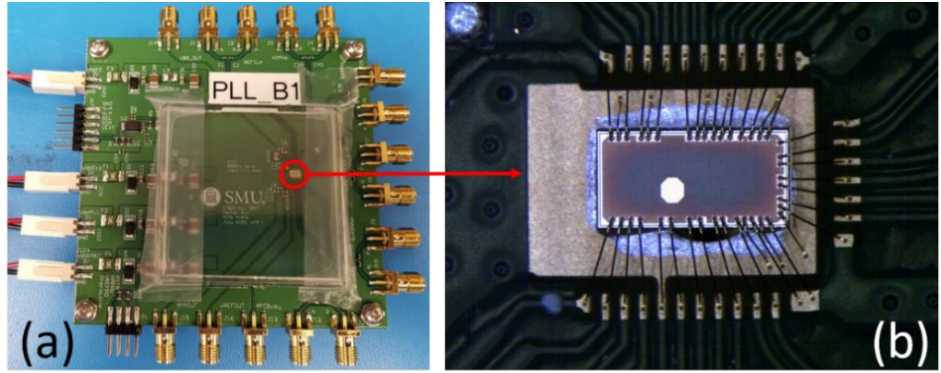

**Figure 3:** Test board (a) and chip photograph (b) of ETROC PLL.

For characterization of ETROC PLL, Lab tests were conducted with a 40 MHz reference clock provided by a Silicon Labs Si5338 crystal clock generator. After AFC calibration, the PLL is locked at 5.12 GHz with the capacitor bank code ranging from 19 to 21. The calibration results on different test boards show great consistency, as shown in figure 4.

The jitter performance of all output clocks has been extensively studied. The ETROC PLL displays a random jitter below 2 ps (RMS) on all output clocks. The Time Interval Error (TIE) jitter of ETROC PLL is measured to be within ±5 ps (peak-to-peak), better than the measured result (±15 ps) of ljCDR in lpGBT v0 [3], as shown in figure 5(a). The jitter improvement could originate from the suppressed power supply noise. In lpGBT v0, the feedback divider in the PLL loop, the clock distribution network that brings the clock signals to the IO pads, and other digital blocks (serializer, etc.) outside the PLL share a core power supply, which is modulated by the 40 MHz clock activity at the chip level. In contrast, in the ETROC PLL standalone chip, the core power supply has no other digital blocks to share and has plenty of decoupling capacitors. During operation, the power consumption of the PLL core is about 61 mW, consisting of 49 mW for the analog parts and 12 mW for the digital parts. Certain digital blocks (the feedback divider, AFC, etc.) are implemented with TMR, which triples the power consumption. The total power consumption is dominated by the analog parts and is not significantly impacted by the radiation hardness technique. All the measured performances meet the ETROC requirements.

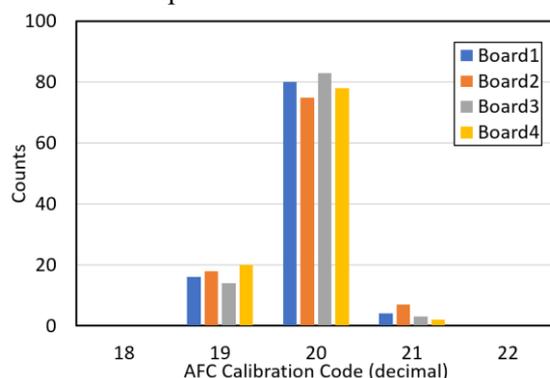

**Figure 4:** AFC calibration code on different test boards.

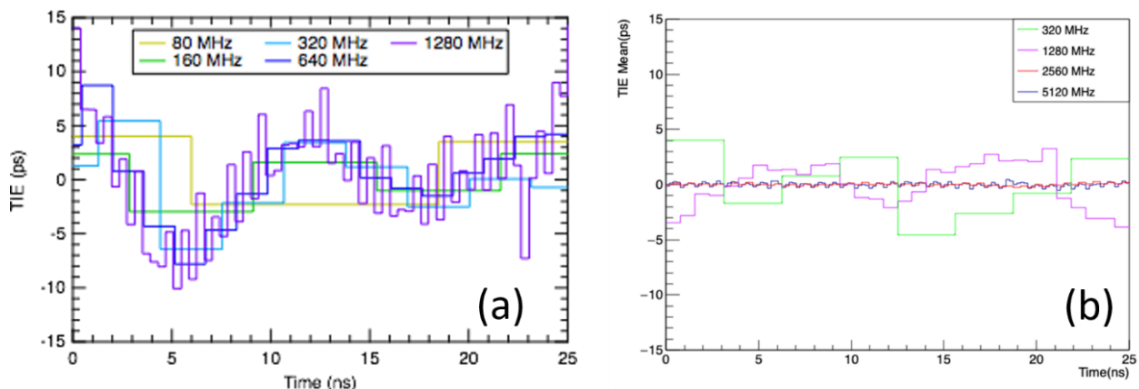

**Figure 5:** TIE jitter of ljCDR in lpGBT v0 (a) and ETROC PLL (b).

## 4. Radiation Test Results

To evaluate the TID tolerance, ETROC PLL was exposed in the X-ray facility with a high dose rate at KU Leuven University, Belgium. ETROC PLL was able to operate without significant degradation when TID reached 300 Mrad.

ETROC PLL test chip was also tested at the Heavy Ion facility in Louvain, Belgium. Heavy Ion irradiation was performed with Linear energy transfer (LET) between 1.3 and 62.5 $MeV \times cm^2/mg$. The fluence per ion was up to $3 \times 10^7/cm^2$. The SEU correction counter in the I²C block behaved as expected. The PLL circuit itself performed stably during irradiation: no unlocks have been identified. The protection function of the AFC TMR register was verified successfully. The AFC calibrated data "Capsel" was protected correctly; thus, no large phase or frequency jump of the PLL loop was observed. Compared to the old version in lpGBT [5], the update of the ljCDR did not appear to introduce extra SEE sensitivity.

The SEE sensitivity outside the PLL loop was observed in the heavy ion irradiation campaign. Figure 6 presents the random sample of positive phase jumps. The short phase jumps with a magnitude between 50 and 600 ps persists for 1 to 3 μs. The saturation cross-section is about $10^{-6}$ cm$^2$. The additional sensitivity could originate in the output clock distribution or IO CML driver biasing. It is possible to identify the origins in a two-photon laser campaign conclusively. A proton beam test is planned to further check the SEEs in the environment where ETROC will operate.

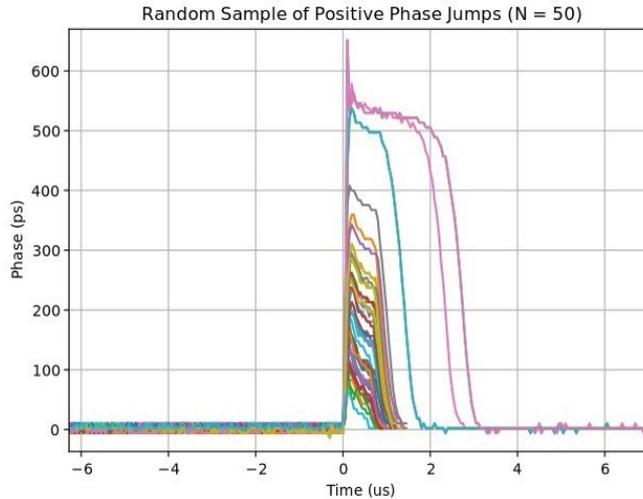

**Figure 6:** Random sample of positive phase jumps during heavy ion irradiations.

## 5. Conclusion and Outlook

The ETROC PLL, based on the ljCDR in lpGBT, has been prototyped and extensively tested. PLL loop locks after the automatic frequency calibration. The power consumption of the PLL core is about 61 mW with a random jitter below 2 ps (RMS). Radiation tolerance of ETROC PLL has been extensively tested both in terms of TID and SEU. No performance degradation was observed with a dose of 300 Mrad. During the SEU testing with heavy ions, the sensitivity of the PLL core for lpGBT v0 remains valid in ETROC PLL, and the protection function of the AFC TMR register was also verified successfully. ETROC PLL meets the design requirements for clock generation in the ETROC and will be integrated into the next iteration chip, ETROC2.

## Acknowledgments

This work has been authored by Fermi Research Alliance, LLC under Contract No. DE-AC02-07CH11359 with the US Department of Energy, Office of Science, Office of High Energy Physics.